\documentstyle[multicol,epsf,aps,pre,twoside]{revtex} 
  \def\sn{\mathop{\rm sn}\nolimits}

  \def\order{\mathop{\rm O}\nolimits}

\begin{document}

\title{\ \\Magnetic critical behavior of  two-dimensional\\
 	random-bond Potts ferromagnets in confined geometries}

\author{Christophe Chatelain and Bertrand Berche\cite{byline2}}
\address{Laboratoire de Physique des Mat\'eriaux,~\cite{byline3} 
Universit\'e Henri Poincar\'e,\\ Nancy
1, BP 239,
F-54506  Vand\oe uvre les Nancy Cedex, France}

\date{23 March 1999}

\maketitle

\begin{abstract}
We present a numerical study of $2D$ random-bond Potts ferromagnets. The model
is studied both below and above the critical value $Q_c=4$ 
which discriminates between second and first-order transitions 
in the pure system. Two geometries
are considered, namely cylinders and square-shaped systems, and the critical
behavior is investigated through conformal invariance techniques which were 
recently shown to be valid, even in the randomness-induced second-order
phase transition regime $Q>4$.
In the cylinder geometry, connectivity transfer matrix calculations provide
a simple test to find the range of disorder amplitudes which is characteristic of
the disordered fixed point. The scaling dimensions then follow from the exponential
decay of correlations along the strip.
Monte Carlo simulations of spin systems on the other hand 
are generally performed on systems of
rectangular shape on the square lattice, but the data are then perturbed
by strong surface effects. The conformal mapping of a semi-infinite 
system inside a square 
enables us to take into account boundary effects explicitly and leads to an accurate
determination of the scaling dimensions. The techniques are applied to different
values of $Q$ in the range 3--64.
\end{abstract}


\begin{multicols}{2}
\narrowtext

\section{Introduction}
\label{sec:sec1}

The presence of impurities can have significative effects on the nature of phase 
transitions. Both from experimental and theoretical perspectives, the study of
the influence of randomness is of great importance. Experimental evidences
of the effect of random quenched impurities in two-dimensional systems were 
found in order-disorder phase transitions of adsorbed atomic layers belonging, in the pure case, to
the $Q=4$-state Potts model universality 
class~\cite{schwengerbuddevogespfnur94,vogespfnur98}. In the presence
of disorder, the critical exponents are modified. On the other hand, no
modification is found when the pure system belongs to the Ising universality
class~\cite{mohankronmullerkelsch98}.  

The study of disordered systems is a quite active field of research in
statistical physics, and the resort to large-scale
Monte Carlo simulations is often helpful~\cite{selkeshchurtalapov94}.
Numerical investigations of the critical properties of random
systems require averages over disorder realizations.
Standard techniques, like Finite-Size Scaling (hereafter referred to as FSS) or temperature dependence 
of the physical quantities were extensively used, and, more recently, 
conformal invariance techniques were shown to provide accurate results.

The effect of quenched bond randomness in a system which undergoes 
a second-order phase transition in the homogeneous case has been considered 
first. It
is well understood since Harris proposed a relevance criterion for the case of
fluctuating interactions~\cite{harris74}.
Disorder appears to be a relevant perturbation  when the specific heat exponent
$\alpha$ of the pure system is positive. Since in the two-dimensional Ising
model (IM) $\alpha$ vanishes due to the logarithmic Onsager singularity, this
model was carefully studied in the 1980s~\cite{shalaev94}. 
The analogous situation when the pure system exhibits a first-order transition
was less well studied, in spite of the early work of Imry and Wortis who
argued that quenched disorder could induce a second-order phase 
transition~\cite{imrywortis79}. This argument was then rigorously proved
by Aizenman and Wehr, and Hui and 
Berker~\cite{aizenmanwehr89,huiberker89}. In two dimensions, even an 
infinitesimal amount of
quenched impurities changes the transition into a continuous one.

The first intensive Monte Carlo (MC) study of  the effect of disorder 
at a first-order 
phase transition is due to Chen, Ferrenberg and 
Landau.
These authors studied the $Q=8$-state two-dimensional 
random-bond Potts model (RBPM), which, 
 in the pure case,  is known to exhibit a first-order phase 
transition when $Q>4$, the larger the value of $Q$, the sharper the 
transition~\cite{wu82}. Taking advantage of duality, they performed
a finite-size scaling study at the critical point of a 
self-dual disordered 
system~\cite{chenferrenberglandau92,chenferrenberglandau95}
and definitively showed that the transition  becomes of
second order in the presence of bond randomness.
Their results, together with other related 
works~\cite{novotnylandau81,kardaretal95,wisemandomany95,cardy96b}, suggested that any 
two-dimensional random system should belong to the 2D pure IM universality 
class. These results were also coherent with real 
experiments~\cite{schwengerbuddevogespfnur94}. 

In recent papers, Cardy and Jacobsen used a different 
approach~\cite{cardyjacobsen97,jacobsencardy98}, based on the 
connectivity transfer matrix (TM)
formalism of Bl\"ote and Nightingale~\cite{blotenightingale82}.
They studied random-bond Potts models for different values of $Q$ and with a bimodal
probability distribution of coupling strengths. Their 
estimations
of the critical exponents  lead to a continuous variation of $\beta /\nu$ 
with $Q$. This result is in accordance with previous theoretical calculations 
and MC simulations when $Q\leq 4$~\cite{dotsenkopiccopujol95a,picco96}.
In the randomness-induced second-order
phase transition regime $Q>4$, $\beta/\nu$ is quite different from the Ising
value of ${1\over 8}$ and particularly in sharp disagreement with the Monte Carlo 
results of Ref.~\cite{chenferrenberglandau95} for $Q=8$. Since then, 
 Monte Carlo simulations
were performed by different groups at 
$Q=8$~\cite{chatelainberche98,yasar98,picco98}.
The choice of the value $Q=8$ was motivated by the value of the correlation 
length in the pure case ($\xi= 23.87$ in lattice spacing units)~\cite{buffenoirwallon93}.
MC simulations which enable to discriminate between a
first-order regime and a second-order transition can indeed be performed
easily with systems of larger sizes. These studies led
to partially 
conflicting results given in Table~\ref{tabcflcjcbp}, but they eventually
found an explanation in terms of a crossover behavior in a recent 
work of Picco~\cite{picco98}. While theoretical calculations are generally
managed in the weak disorder regime (perturbation expansion around the 
homogeneous system fixed point), the range of disorder amplitude must be
chosen carefully in numerical studies, since the random fixed point (FP) can
be perturbed by crossover effects due to the pure and/or the percolation unstable
fixed points. The disordered FP properties are thus more easily observed with
strong randomness.
A disorder amplitude $r$, given by the ratio
of the two types of couplings (distributed according to a binary distribution), 
in the range 8-20 appears to be adapted
to a numerical analysis and gives a good estimate of the disordered fixed point
exponents~\cite{picco98,cardy98} as already observed in the $2D$ random-bond Ising model 
(RBIM)~\cite{andreichenkoetal90c,wangetal90a}. 

\vbox{
\narrowtext
\begin{table}
\caption{Bulk magnetic scaling index obtained by different groups in
the 8-state Potts model.\label{tabcflcjcbp}}
\begin{tabular}{llll}
	Authors & $r$ & $\beta /\nu$ & Technique \\
	\hline
	Chen {\it et al}, Ref.~\protect\(\cite{chenferrenberglandau95}\protect\) & 2 & 0.118(2) & MC \\
	Cardy and Jacobsen, Ref.~\protect\(\cite{cardyjacobsen97}\protect\) & 2 & 0.142(4) & TM \\
	Chatelain and Berche, Ref.~\protect\(\cite{chatelainberche98}\protect\)  & 10 & 0.153(3) & MC \\
	Picco, Ref.~\protect\(\cite{picco98}\protect\)  & 10 & 0.153(1) & MC \\
\end{tabular}
\end{table}
}

The surface properties of dilute or random-bond magnetic systems were paid less attention. 
The whole set of bulk and surface critical exponents of a given system is 
determined by the anomalous dimensions of the relevant scaling fields which 
enter the homogeneity assumption of the singular free energies~\cite{binder83}.
The
$(1,1)$ surface of the disordered Ising model on a square lattice has only recently been 
investigated through MC simulations by Selke {\it et al.}~\cite{selkeetal97}.
The critical exponent $\beta_1$ of the boundary magnetization was found to be 
equal within error bars to its value in the pure 2D IM. The surface properties of the 
8-state RBPM were also computed in Ref.~\cite{chatelainberche98}.

In this paper, we are interested in the bulk critical behavior
of disordered Potts ferromagnets, and in the evolution of their properties
as the number of states $Q$ increases. The Hamiltonian of the model is
\begin{equation}
   -\beta{\cal H}=\sum_{(ij)}K_{ij}\delta_{\sigma_i,\sigma_j}
   \label{Ham}
\end{equation} 
where the spins can take $Q$ different values
and the coupling strengths between nearest neighbor spins 
are taken from a binary probability
distribution
\begin{equation}
{\cal P}(K_{ij})=p\delta(K_{ij}-rK)+(1-p)\delta(K_{ij}-K) 
   \label{probcoupl}
\end{equation}
with $p=1/2$, which guarantees the self-duality relation
\begin{equation}
(e^{rK_c}-1)(e^{K_c}-1)=Q.   
   \label{selfdual}
\end{equation}
The value $r=1$ corresponds to the pure model and $r\to \infty$ to the
percolation limit.

In the present work, following previous studies, 
we use the powerful methods of conformal invariance.
 Talapov {\it et al.}  studied
numerically the critical-point correlation functions in the $2D$ RBIM 
on the torus~\cite{difrancesco87} and took  into 
account the finite-size effects through
a convenient conformal rescaling~\cite{talapovdotsenko93,talapovshchur94}.
In the cylinder geometry,  conformal invariance methods have
also been successfully applied. In the two-dimensional RBIM, 
randomness being a marginally irrelevant perturbation, many results
have been obtained {\it via} these techniques: Conformal anomaly, correlation 
decay, gap-exponent relation for long strips~\cite{dQS94,dQ95,AdQdS96}.
At randomness-induced second-order phase transitions, conformal
techniques have also been used already~\cite{cardyjacobsen97,jacobsencardy98,picco97}
and numerical evidences for the validity of the conformal covariance
assumption for correlation functions and density profiles were recently 
reported~\cite{chatelainberche98b}.
It is well known that in disordered spin systems, the strong fluctuations
of couplings from
sample to sample require careful averaging 
procedures~\cite{derrida84,wisemandomany98a,wisemandomany98}. For 
that reason, the study of the probability distributions must be performed
in order to guarantee that the average quantities, which should obey the
conformal covariance assumption, are correctly obtained numerically. 
A comparison between {\it grand canonical disorder} (GCD) and 
{\it canonical disorder} (CD)
will also be given.


The plan of the paper is the following: In Sec~\ref{sec:sec2}, we
present the results of connectivity transfer matrix calculations on strips
with periodic boundary conditions for different values of $Q$.
The order parameter correlation function, after disorder average, leads
to estimates of the magnetic scaling index for different strip sizes.
From our knowledge in the case $Q=8$~\cite{chatelainberche98b}, it appears that these computations
are suitable for the determination of a convenient disorder amplitude in
order to reach the disordered FP. At large disorder amplitudes
($r\simeq 10$), the behavior of the effective central charge can indeed
discriminate between random and percolation fixed points. In Sec.~\ref{sec:sec3}, we 
report Monte Carlo simulations in a square geometry with the above-mentioned
disorder amplitude. The 
magnetization correlation function and density profile give access to refined
values for the corresponding exponents. A discussion
of the results is given in Section~\ref{sec:sec4}.
Attention is paid to take into
account the different sources of error for the results reported in this work.


\section{Cylinder geometry and disordered fixed point}\label{sec:sec2}

\subsection{Free energy and central charge}

In the strip geometry, we used the Bl\"ote and Nightingale connectivity 
transfer matrix method~\cite{blotenightingale82}. In disordered systems, transfer operators
in the time direction do not commute and, as a consequence, the free energy density
is no longer defined by the largest eigenvalue of a single TM, but in terms of the 
leading Lyapunov exponent.
For a strip of size $L$
with periodic boundary conditions, the 
leading Lyapunov exponent follows from the Furstenberg method~\cite{furstenberg63}:
\begin{eqnarray}
	\Lambda_0(L)&=&\lim_{m\to\infty}\frac{1}{m}
	\ln\left|\!\left|\left(\prod_{k=1}^m 
	{\bf T}_k\right)
	\mid\! v_0\rangle\right|\!\right|\nonumber\\
	&=&\lim_{m\to\infty}\Lambda_0(L;m),
\label{eq-Furst}
\end{eqnarray}	
where ${\bf T}_k$ is the transfer matrix between columns $k-1$ and $k$ and
$\mid\! v_0\rangle$ is a suitable unit initial vector. The  
free energy density
is thus given by
\begin{equation}
	[f_0(L)]_{\rm av}=-L^{-1}\Lambda_0(L),
	\label{eq-f0}
\end{equation}	
where $[\dots]_{\rm av}$ denotes the average over disorder realizations.

\begin{figure}[h]
	\epsfxsize=8cm
	\begin{center}
	\mbox{\epsfbox{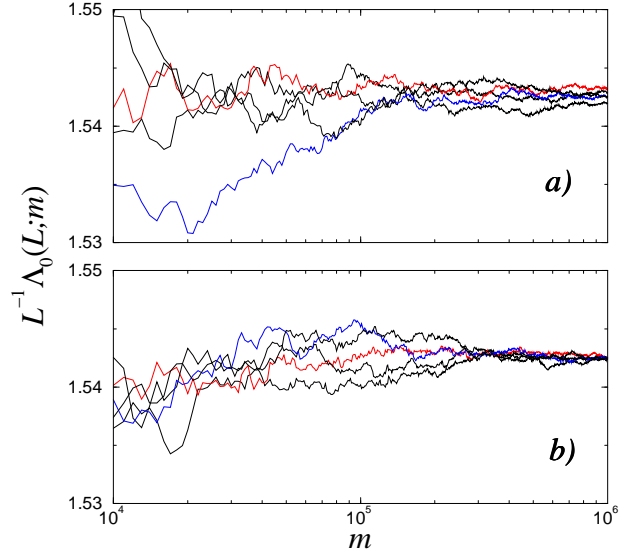}}
	\end{center}
	\vskip -0.5cm
	\caption{Free energy density (up to an additive constant $\ln Q$) 
	{\it vs} $m$, the number of iterations of the TM for a strip of size 
	$L=6$  ($Q=8$, $r=10$) with 5 realizations in grand canonical (a) and
	canonical (b) disorder.}
	\label{GCan-Can}  
\end{figure}

In the following, we considered
canonical disorder, a situation in which
 exactly the same numbers of  couplings $K$ and $rK$ are 
distributed over the
bonds of the whole system of length $\sim 10^6$. This choice contributes to reduce sample
fluctuations. This is shown in Fig.~\ref{GCan-Can} where the stability of
the free energy density is compared
to the standard grand canonical disorder
for different runs up to $m=10^6$ iterations of the TM.

In Eq.~(\ref{eq-Furst}), the disorder average is implicitly performed through
an infinite number of iterations of the transfer matrix. In our computations,
only a finite number $m$ is used, leading to approximate values denoted by
$\Lambda_0^{(i)}(L;m)$ for different runs
labelled by an integer $i=1,M$. 
The leading Lyapunov exponent and the corresponding eigenvector, 
$\mid\!\Lambda_0\rangle$, obtained after 
$m=10^6$ iterations of the TM, are then averaged over $M=48$ independent runs. 
The average free energy density of 
Eq.~(\ref{eq-f0}) is thus replaced, in the calculations, by
\begin{eqnarray}
	[f_0(L)]_{\rm av}&\simeq& [f_0(L)]_M\nonumber\\
	&=&-L^{-1}\left(\frac{1}{M}\sum_{i=1}^M
	\Lambda_0^{(i)}(L;10^6)\right).
\label{eq-f0bis}
\end{eqnarray}	
The value $M=48$ was chosen in order to guarantee a stability of the averaged 
quantities with a relative error better than $10^{-5}$ for the free 
energy density and than $4\times
10^{-5}$ for the components of the corresponding eigenvector. 
The  computations are then 
performed on strips of sizes $L=2$ to 8.

\begin{figure}[h]
	\epsfxsize=8cm
	\begin{center}
	\mbox{\epsfbox{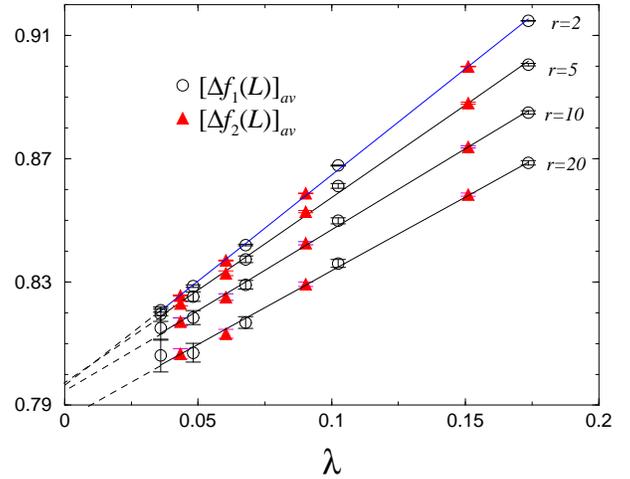}}
	\end{center}
	\vskip -0.5cm
	\caption{Reduced difference between the free energies at 
	different sizes $[\Delta f_l(L)]_{\rm av}$ for different values of the
	disorder amplitude $r$ ($Q=3$). The central charge is
	given by the intercept {\it via} a
	linear fit. The parameter $l$ is defined in 
	Eq.~\protect\( (\ref{eq-deltaf})\protect\).}
	\label{Deltaf-lambda}  
\end{figure}

The numerical investigation of critical properties in random systems 
requires the knowledge of the range of disorder amplitude (measured here by the 
ratio $r$
between strong and weak couplings) for which the  fixed point properties is 
reached. Outside this regime, strong crossover effects perturb the 
data~\cite{chatelainberche98b}.
A convenient disorder amplitude $r$ can be obtained
from the behavior of the effective central charge, which
{\it increases} when the system approaches the disordered fixed 
point in non-unitary 
theories as it seems to be the case in the RBPM~\cite{jacobsencardy98,dotsenkojacobsenlewispicco98}.
The central charge $c$ is defined by the leading size dependence of
the free energy density, and, since the strip sizes are quite small,
corrections to scaling must be included:
\begin{equation}
	[f_0(L)]_{\rm av}=f_{\rm reg}-\frac{\pi c}{6}L^{-2}+AL^{-4}.
\label{eq-corr-f}
\end{equation}	
The comparison between successive sizes $L$ and $L+l$ allows us to define a reduced 
difference which leads to
\begin{eqnarray}
	[\Delta f_l(L)]_{\rm av}&\equiv& \frac{6}{\pi}\frac{[f_0(L)]_{\rm av}-[f_0(L+l)]_{\rm av}}
	{(L+l)^{-2}-L^{-2}}\nonumber\\
	&=&
	c-\frac{6}{\pi}A\lambda ,
	\label{eq-deltaf}
\end{eqnarray}	
where the reduced parameter $\lambda$ is given by
\begin{equation} 
	 \lambda=\frac{(L+l)^{-4}-L^{-4}}{(L+l)^{-2}-L^{-2}}. 
\label{eq-lambda}
\end{equation}	
In the thermodynamic
limit, the central charge $c$ then
follows from a linear fit as shown in Fig~\ref{Deltaf-lambda}
for strips of sizes $L=2$ to 8 in the case $Q=3$.  We restricted our study to integer values of $r$ and the data for
the effective central charge at different disorder amplitudes are
given in the Table~\ref{tab-c}. We observe that 
the value of $c$ is
strongly depending on the disorder amplitude: It increases from
the weak disorder limit up to a maximum value and then decreases slowly as
$r$ increases.

\begin{figure}[h]
	\epsfxsize=8cm
	\begin{center}
	\mbox{\epsfbox{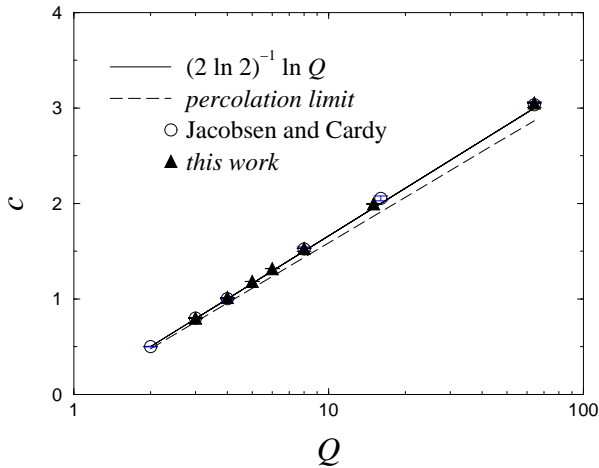}}
	\end{center}
	\vskip -0.5cm
	\caption{Central charge at the random fixed point as a function of the number of states. The full
	line corresponds to $c(Q)=\ln Q/(2\ln 2)$ while the dashed line is the
	percolation limit $c(Q)=5\sqrt 3\ln Q / 4\pi$. Error bars are smaller
	than the sizes of the symbols.}
	\label{c-vs-Q}  
\end{figure}

The central charge at the random fixed point ({\it i.e.} the maximal value obtained
for an {\it optimal disorder amplitude} $r^\star(Q)$)
is shown in Fig.~\ref{c-vs-Q}. 
Assuming a
 linear behavior in $\ln Q$~\cite{picco97} which preserves the Ising 
 value $c(Q=2)=1/2$~\cite{jacobsencardy98},
 one gets 
 \begin{equation}
 	c(Q)=\frac{\ln Q}{2\ln 2}
 	\label{eq-c-de-Q}
 \end{equation}
whilst the percolation
 limit leads to $c(Q)=\frac{5\sqrt 3}{4\pi}\ln Q$~\cite{jacobsencardy98}.  
 The two behaviors
 are shown in Fig.~\ref{c-vs-Q}.  The numerical data are in good agreement 
 with Eq.~\ref{eq-c-de-Q} and are accurate enough  to consider
 that the random FP has been reached at $r^\star$ (whose values 
are coherent with those found by Jacobsen and Cardy~\cite{jacobsencardy98}:
$r^\star(3)=7$, $r^\star(8)=9$ and $r^\star(64)=10$).
 In the following, the scaling properties will be
studied at the  optimal  disorder amplitudes in contradistinction with 
previous papers~\cite{cardyjacobsen97,jacobsencardy98}.

\end{multicols}
\widetext
\begin{table}
\caption{Extrapolation of the effective central charge $c$ 
in the thermodynamic
limit for the different 
values of $Q$ and $r$. At each $Q$, the larger values of $c$ 
(written in bold face) 
correspond to the random fixed point regime with an optimal disorder 
amplitude $r^\star$. This maximum is not always located at the same $r$, 
as shown
in the case $Q=3$ and $64$ for two different runs.
\label{tab-c}}
\begin{tabular}{l|lllllllll}
&\multicolumn{9}{c}{Effective central charge at $Q=3$}\\
\tableline
$r$ & 2 & 4 & \bf 5 & \bf 6 & 7 & 10 & 20 &   &  \\ 
$c$ & 0.7970  & 0.7998 & \bf    0.79984 & \bf   0.79969 & 0.7992 & 0.7970 & 0.7879 & &    \\ 
$\Delta c$ & 4$\times 10^{-4}$  & 4$\times 10^{-4}$ & $\bf 3.8\times 10^{-4}$ & $\bf 3.8\times 10^{-4}$ & 4$\times 10^{-4}$ & 4$\times 10^{-4}$ & 4$\times 10^{-4}$ &   &  \\ 
$c$ &   &  &     0.8005 & \bf   0.80070 & \bf 0.80099 &  &  & &    \\ 
$\Delta c$ &   &  & $ 4\times 10^{-4}$ & $\bf 3.8\times 10^{-4}$ &\bf 3.7$\times 10^{-4}$ &  &  &   &  \\ 
\tableline
\tableline
&\multicolumn{9}{c}{Effective central charge at $Q=4$}\\
\tableline
$r$ & 2 & 5 & \bf 6 & \bf 7 & 8 & 10 & 20 &   &  \\ 
$c$ & 1.0043 & 1.0144 & \bf     1.01495 & \bf       1.01483 & 1.0142 & 1.0123 & 0.9996 &   &  \\ 
$\Delta c$ & 4$\times 10^{-4}$ & 4$\times 10^{-4}$ & $\bf 4.3\times 10^{-4}$ & $\bf 4.3\times 10^{-4}$ & 4$\times 10^{-4}$ &4$\times 10^{-4}$  & 4$\times 10^{-4}$ &   &  \\ 
\tableline
\tableline
&\multicolumn{9}{c}{Effective central charge at $Q=5$}\\
\tableline
$r$ & 2 & 5 & 6 & \bf 7 & \bf 8 & 10 & 20 &   &  \\  
$c$ & 1.1579 & 1.1794 & 1.1810 & \bf      1.181593 & \bf   1.181326 & 1.1794 & 1.1642 &    &  \\ 
$\Delta c$ & 5$\times 10^{-4}$ & 5$\times 10^{-4}$ & 5$\times 10^{-4}$ & $\bf 4.6\times 10^{-4}$ & $\bf 4.6\times 10^{-4}$ & 5$\times 10^{-4}$ & 5$\times 10^{-4}$ &    &  \\ 
\tableline
\tableline
&\multicolumn{9}{c}{Effective central charge at $Q=6$}\\
\tableline
$r$ & 2 & 5 & \bf 7 & \bf 8 & \bf 9 & 10 & 20 &   &  \\  
$c$ & 1.2764 & 1.3128 & \bf   1.3172 & \bf      1.3174 & \bf   1.3168 & 1.3157 & 1.2986 &    &  \\ 
$\Delta c$ & 5$\times 10^{-4}$ & 5$\times 10^{-4}$ & $\bf 5\times 10^{-4}$ & $\bf 5\times 10^{-4}$  & $\bf 5\times 10^{-4}$ & 5$\times 10^{-4}$ & 5$\times 10^{-4}$ &    &  \\ 
\tableline
\tableline
&\multicolumn{9}{c}{Effective central charge at $Q=8$}\\
\tableline
$r$ & 2 & 5 & \bf 9 & \bf 10 & \bf 11 & 12 & 20 &   &  \\  
$c$ & 1.4468 & 1.5203 & \bf   1.5329 & \bf      1.5300 & \bf   1.5287 & 1.5270 & 1.5104 &    &  \\ 
$\Delta c$ & 5$\times 10^{-4}$ & 5$\times 10^{-4}$ & $\bf 5\times 10^{-4}$ & $\bf 5\times 10^{-4}$ & $\bf 5\times 10^{-4}$ & 5$\times 10^{-4}$ & 5$\times 10^{-4}$ &    &  \\ 
\tableline
\tableline
&\multicolumn{9}{c}{Effective central charge at $Q=15$}\\
\tableline
$r$ & 2 & 5 & 9 & \bf 10 & \bf 11 & 12 & 13 & 15 & 20 \\ 
$c$ &  1.7313 & 1.9606 & 1.9963 & \bf       1.9937 & \bf    1.9930 & 1.9915 & 1.9895 & 1.9846 & 1.9708\\
$\Delta c$   & 6$\times 10^{-4}$ & 6$\times 10^{-4}$ & 6$\times 10^{-4}$ & $\bf 6\times 10^{-4}$ & $\bf 6\times 10^{-4}$ & 6$\times 10^{-4}$ & 6$\times 10^{-4}$ & 6$\times 10^{-4}$  & 6$\times 10^{-4}$ \\
\tableline
\tableline
&\multicolumn{9}{c}{Effective central charge at $Q=64$}\\
\tableline
$r$ & 2 & 5 & 10 & \bf 11 & \bf 12 & 13 & 15 & 20 &  \\ 
$c$ & 2.0302 & 2.9351 & 3.0414 & \bf    3.0432  &\bf      3.0430   & 3.0415 & 3.0362 & 3.0182 &   \\
$\Delta c$ & 7$\times 10^{-4}$ & 7$\times 10^{-4}$ & 7$\times 10^{-4}$  & $\bf 7\times 10^{-4}$ & $\bf 7\times 10^{-4}$ &7$\times 10^{-4}$    & 7$\times 10^{-4}$  & 7$\times 10^{-4}$ &    \\
$c$ &  &  &  & \bf    3.0526  &\bf      3.0528   & 3.0516 &  &  &   \\
$\Delta c$ &  &  &   & $\bf 7\times 10^{-4}$ & $\bf 7\times 10^{-4}$ &7$\times 10^{-4}$    &   &  &    \\
\end{tabular}
\end{table}
%
\begin{multicols}{2}
\narrowtext

\subsection{Probability distribution of the correlation function}

For a specific disorder realization, the spin-spin correlation function
along the strip 
\begin{equation}
\langle G_{\sigma}(u)\rangle=\frac{Q\langle\delta_{
\sigma_j\sigma_{j+u}}\rangle-1}{Q-1},
\label{eq-Gu}
\end{equation}	
 where $\langle\dots\rangle$ denotes the 
thermal average, is given by the probability that the spins along some row,  
at columns $j$
and $j+u$, are in the same state ($j$
and $j+u$ measure the position in the longitudinal
direction of the strip):	
\begin{equation}
	\langle\delta_{\sigma_j\sigma_{j+u}}\rangle=\frac{
	\langle \Lambda_0\!\mid{\bf g}_j
	\left(\prod_{k=j}^{j+u-1}
	{\bf T}'_k\right){\bf d}_{j+u}\mid\! \Lambda_0\rangle}{
	\langle \Lambda_0\!\mid
	\prod_{k=j}^{j+u-1}
	{\bf T}_k\mid\! \Lambda_0\rangle},
\label{eq-corr}
\end{equation}
where $\mid\! \Lambda_0\rangle$ is the ground state eigenvector and 
${\bf T}_k'$ is the 
transfer matrix in the extended Hilbert space which includes the connectivity
with the origin site $j$. The operator 
${\bf g}_j$ identifies the cluster containing $\sigma_j$, while
${\bf d}_{j+u}$ gives the appropriate weight depending on whether or not
$\sigma_{j+u}$ is in the same state as $\sigma_j$. The computation
is performed
with a grand canonical disorder.

An analysis of the correlation function
probability distribution is needed in order to ensure 
that self-averaging problems do not alter the mean
values~\cite{crisantinicolispaladinvulpiani90}. 
The methodology that we propose is to deduce the critical behavior from the decay of the
correlation functions using conformal symmetry.
Since conformal covariance assumption is supposed
to be satisfied by {\it average quantities}, i.e. 
$[\langle{G}_{\sigma}(u)\rangle]_{\rm av}$, our first
aim is to show that, in spite of the lack of self-averaging, our numerical
experiments lead to well-defined
averages. 

\begin{figure}
	\epsfxsize=8cm
	\begin{center}
	\mbox{\epsfbox{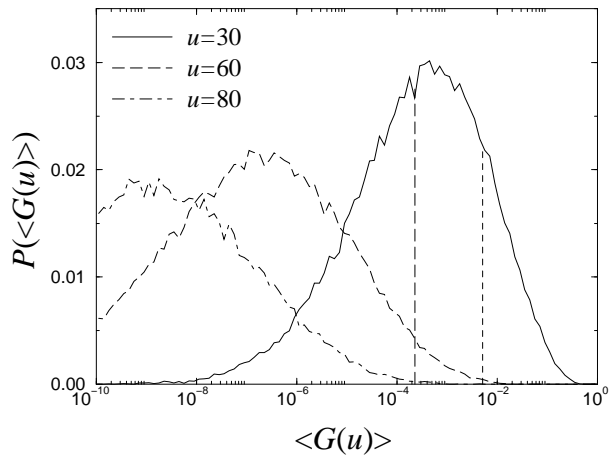}}
	\end{center}
	\caption{Probability distribution of the correlation function after
	63436 realizations of disorder for a strip of size $L=6$ 
	($Q=8$, $r=10$). The vertical dotted line shows the average value 
	$[\langle G_\sigma(30)\rangle]_{\rm av}$ while the long-dashed line
	shows the typical value ${\rm e}^{[\langle \ln G_\sigma(30)\rangle]_{\rm av}}$.
	}
	\label{fig-4}  
\end{figure}

The probability distribution of the correlation 
function, as shown in Fig.~\ref{fig-4}, enables us to determine 
the most probable  (or typical) value
$G_\sigma^{mp}(u)$ and the average correlation function 
$[\langle{G}_\sigma(u)\rangle]_{\rm av}$, as well as the
averaged logarithm, $[{\ln\langle G}_\sigma(u)\rangle]_{\rm av}$ at any value
of the distance $u$. Compatible behaviors are found for $G_\sigma^{mp}(u)$ and 
$e^{[{\ln\langle G}_\sigma(u)\rangle]_{\rm av}}$. It is a confirmation of 
the essentially log-normal 
character of the probability
distribution~\cite{crisantinicolispaladinvulpiani90}, as argued by 
Cardy and Jacobsen~\cite{cardyjacobsen97}. 
It is thus necessary to perform averages over 
larger numbers
of samples for $[\langle{G}_\sigma(u)\rangle]_{\rm av}$ than for 
$[{\ln\langle G}_\sigma(u)\rangle]_{\rm av}$ to get the same relative errors.

Following Cardy and Jacobsen, since the moments of the logarithm of the correlation
function are self-averaging,
a cumulant expansion can then be
performed to reconstruct $[\langle{G}_\sigma(u)\rangle]_{\rm av}$ and
to compare to the values obtained by averaging directly over the samples.

\begin{figure}
	\epsfxsize=8cm
	\begin{center}
	\mbox{\epsfbox{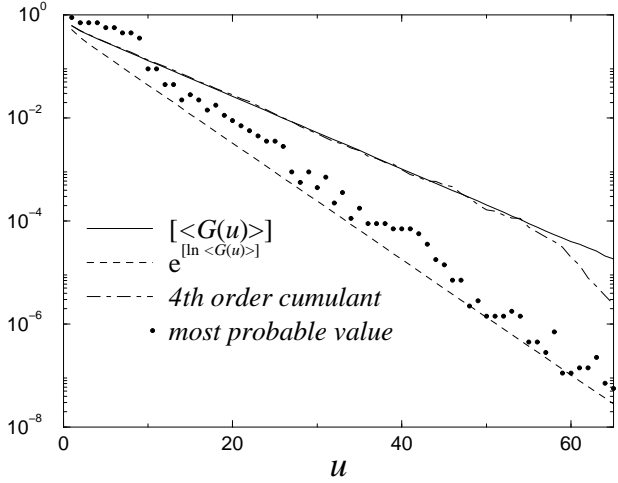}}
	\end{center}
	\caption{Average correlation function, most probable  (or typical) value and sum up to 
	the
	4th order of the 
	cumulant expansion obtained from 63436 
	realizations of disorder for a strip of size $L=6$ ($Q=8$, $r=10$).}
	\label{fig-5}  
\end{figure}

The results in Fig.~\ref{fig-5} (for $Q=8$), strengthen the credibility
of the direct average, and also clearly show that the 
cumulant expansion up to fourth order still strongly fluctuates 
at large distances compared to 
$[\langle{G}_\sigma(u)\rangle]_{\rm av}$. In the following we will thus favour the
direct averaging process, using a large number of disorder 
realizations.

\subsection{Bulk magnetic scaling dimension}

We will now use the results that follow from the assumption of
conformal covariance of the average correlation functions. In the infinite
complex plane $z=x+{\rm i}y$ (denoted by the index $\infty$) 
  the correlation function exhibits the usual
algebraic decay at the critical point
\begin{eqnarray}
	[\langle G_\sigma(R)\rangle]_{\rm av}&\equiv&
	[\langle \sigma(z_1)\sigma(z_2)\rangle_{\infty} ]_{\rm av}\nonumber\\
	&=&{\rm const}
	\times 
	R^{-2x_\sigma^b}, 
	\label{eqinfty}
\end{eqnarray}	
where
$R=\mid\! z_1-z_2\!\mid$ and $x_\sigma^b=\beta /\nu$ is the bulk magnetic scaling 
dimension. Under a conformal mapping $w(z)$, the correlation functions of
a conformally invariant 2D-system transforms into the new geometry 
according to
\begin{equation}
	{G}_\sigma(w_1,w_2)=\mid\! w'(z_1)\!\mid^{-x_\sigma^b}
	\mid\! w'(z_2)\!\mid^{-x_\sigma^b} {  G}_\sigma(z_1,z_2).
	\label{eq-transfcorr}
\end{equation}
The logarithmic tranformation $w=\frac{L}{2\pi}\ln z$ is known to map the
$z$ plane onto an infinite strip (denoted by the index st) $w=u+{\rm i}v$ of width $L$ with periodic 
boundary conditions in the transverse direction.
Applying Eq.~(\ref{eq-transfcorr}) in the random system where
$[\langle G_\sigma(w_1,w_2)\rangle ]_{\rm av}\equiv
[\langle \sigma(w_1)\sigma(w_2)\rangle_{\rm st} ]_{\rm av}$ corresponds to the
strip geometry, one gets the usual 
exponential decay along the 
strip
\begin{equation} 
[\langle{G}_\sigma(u)\rangle]_{\rm av}={\rm const}\times
\exp\left(-\frac{2\pi}{L}x_\sigma^b u\right),
\label{eq-expdecay}
\end{equation} 
where $u={\rm Re}\ (w_2-w_1)$. The scaling 
index $x_\sigma^b$ can thus be deduced from a linear fit in a semilog plot.

\begin{figure}
	\epsfxsize=8cm
	\begin{center}
	\mbox{\epsfbox{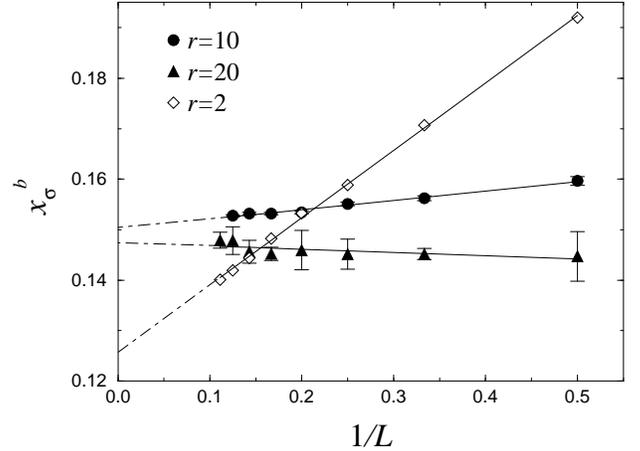}}
	\end{center}
	\caption{Magnetic scaling index deduced from the algebraic decay of 
	the average correlation function along
	the strip of size $L$ as a function of $L^{-1}$ and extrapolation in
	the thermodynamic limit ($Q=8$, $L=2-9$ for $r=2$ and 20, from 
	Ref.~[38]
	and $L=2-8$ for $r=10$, this work, where the data analysis is 
	more refined (see Appendix A) leading to error bars 10 times smaller).}
	\label{fig-Q8}
\end{figure}

For each strip size ($L=2-8$), we realized $80\times 10^3$ disorder 
configurations. 
 It allowed us to define 
 mean values and error bars for the correlation functions at any point in the
 range $u=1-100$, taking into
 account the standard deviation over the samples. The non 
 self-averaging behavior of the
 correlation functions induces large variances (The reduced
 variance 
 $R_X(L)\equiv {([X^2]_{\rm av}-[X]_{\rm av}^2)}/{[X]_{\rm av}^2}$ does not
 behave as a power law, but
 evolves towards a constant value when the strip size increases, {\it e.g.}
 $R_{G_\sigma(20)}(L)\to 1.50$,
 as already observed for several quantities by Wiseman and Domany
 in Refs.~\cite{wisemandomany98a,wisemandomany98}.).
The exponents follow from an 
exponential fit in the range 
$u>5$ and $[\langle{G}_\sigma(u)\rangle]_{\rm av}>\epsilon$, where the cutoff
$\epsilon$ is introduced in order to avoid tiny numbers whose values are
lower than the fluctuations.
The error bars given for the exponents take into account 
the uncertainties of data for the correlation functions~\cite{pressetal92}.
The resulting values for each strip size are plotted against $L^{-1}$ which 
allows an extrapolation in the thermodynamic limit. This is shown in 
Fig.~\ref{fig-Q8} in 
the case $Q=8$. 
 This figure provides a confirmation of the effect of a too weak disorder: 
 Strong crossover effects take place  which lead to a wrong  determination
 of the critical behavior with the strip sizes used here. On the other hand, 
 at the optimal value $r^\star(Q)$,
 the exponent
converges in the $L\to\infty$ limit  towards a
well defined final
estimate.

\begin{figure}
	\epsfxsize=8cm
	\begin{center}
	\mbox{\epsfbox{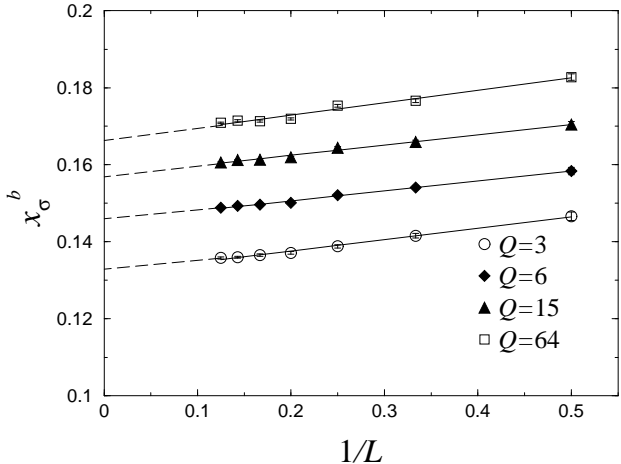}}
	\end{center}
	\caption{Magnetic scaling index deduced from the algebraic decay of 
	the average correlation function along
	the strip of size $L$ as a function of $L^{-1}$ and extrapolation in
	the thermodynamic limit for different $Q$--values ($L=2-8$). }
	\label{fig-allQ}  
\end{figure}

  The convergence of effective scaling dimensions at 
 different
 strip sizes, obtained with a
 cutoff value in the range    $\epsilon=10^{-4}-10^{-6}$ and $r=r^\star(Q)$, is shown in Fig.~\ref{fig-allQ} 
 for different values of $Q$.
 The extrapolation in the thermodynamic limit
 is given in Table~\ref{tabxmG}. The details of the fitting procedure
 and of the evaluation of errors is presented in Appendix~\ref{sec:appendixA}.

\vbox{
\begin{table}
\caption{Bulk magnetic scaling index (after extrapolation in the 
thermodynamic limit) obtained from the decay of the correlation
function along the strip (cutoff parameter $\epsilon=10^{-4}-10^{-6}$).\label{tabxmG}}
\begin{tabular}{llll}
	 $Q$ & $r$ & $x_\sigma^b$ & $\Delta x_\sigma^b$   \\
	\hline
	$\vphantom{10^{2^2}}$3 & 5 & 0.1321 & $3\times 10^{-4}$ \\
	4 & 7 & 0.1385 & $3\times 10^{-4}$  \\
	5 & 7 & 0.1423 & $3\times 10^{-4}$  \\
	6 & 8 & 0.1456 & $3\times 10^{-4}$  \\
	8 & 10 & 0.1505 & $3\times 10^{-4}$  \\
	15 & 10 & 0.1572 & $3\times 10^{-4}$  \\
	64 & 12 & 0.1669 & $3\times 10^{-4}$  \\
\end{tabular}
\end{table}
}


\section{Square geometry and critical behavior}\label{sec:sec3}

\subsection{Conformal rescaling of boundary effects}

Monte Carlo simulations of two-dimensional spin systems are 
generally performed on systems of square shape.  
In the following, we consider such a  system of size
$N\times N$, and call $u$ and $v$ the corresponding directions 
(Fig.~\ref{fig: 1}).

\vspace{-0.0cm}
\vbox{
\begin{figure} [h]
	a)\vspace{-0mm}
	\epsfxsize=11.8cm
	\begin{center}
	\mbox{\qquad\qquad\epsfbox{3dCorr.eps}}
	\end{center}
	\vspace{-80mm}
\end{figure}
}\vbox{
\begin{figure} [h,t]
	b)\vspace{-8mm}
	\epsfxsize=5cm
	\begin{center}
	\mbox{\epsfbox{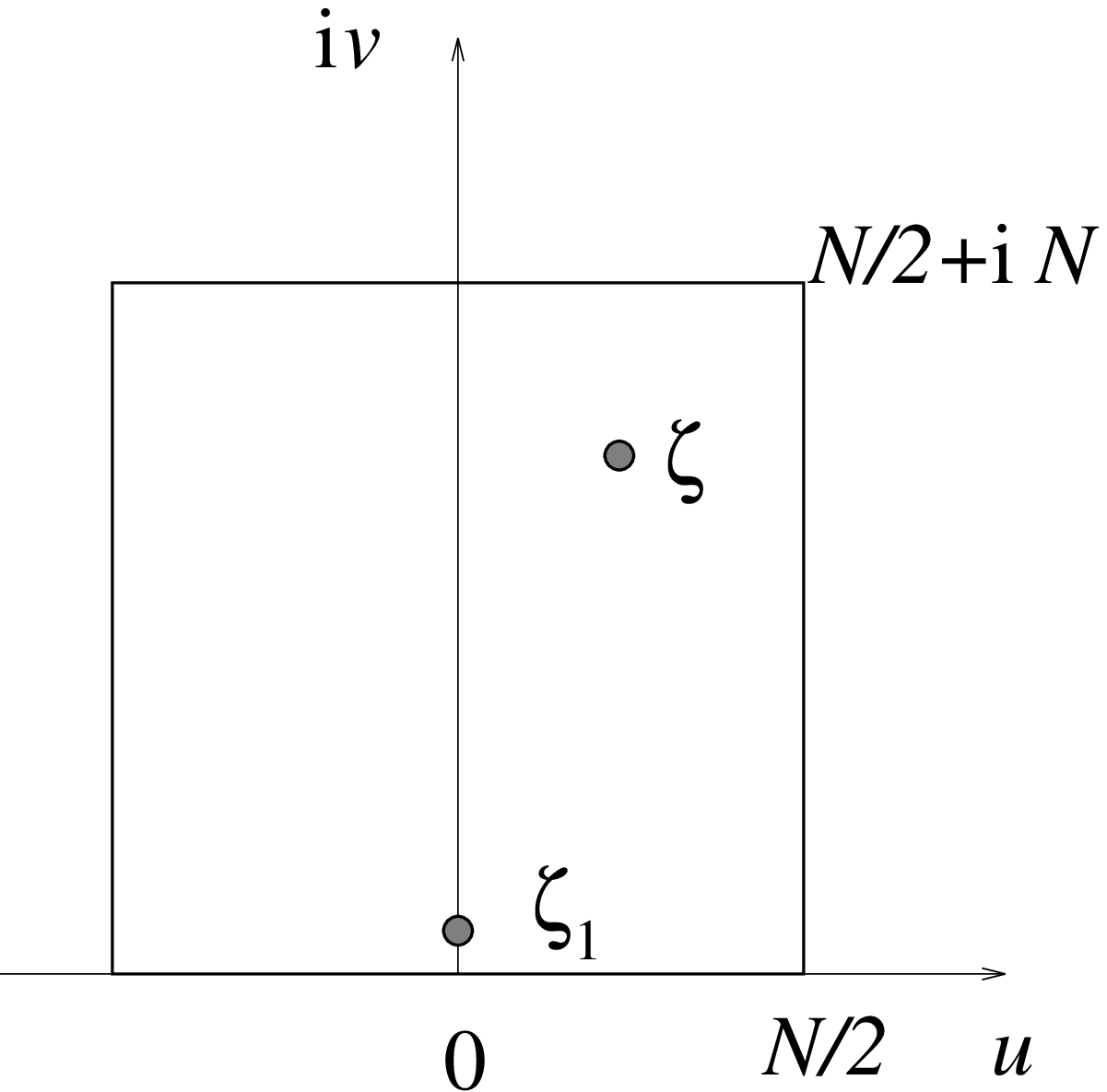}}
	\end{center}
	\vskip 0.3cm
	\caption{a) Monte Carlo simulations of the 2$d$ RBPM inside 
	a square of $101\times 101$ lattice sites
	($10^6$ MCS/spin, Swendsen-Wang cluster algorithm). The 
	figure shows the correlation function between a point
	close to the surface ($\zeta_1={\rm i}$) and all other points $\zeta$ in the 
	square. The notations are specified in b).}
	\label{fig: 1}  \vskip -0cm
\end{figure}
}

The order parameter correlation function between a point close to the surface, and a
point in the bulk of the system should, in principle, lead to both surface and
bulk critical exponents, possibly to structure constants~\cite{barkemamccabe96}. 
Practically, FSS techniques are not of great help for the accurate determination of critical exponents,
since 

i) strong surface effects (shape effects) occur which modify the large distance power-law 
behavior, {\it i.e.}  the scaling regime,

ii) the universal scaling function entering the correlation 
function is likely to display a crossover before its asymptotic 
regime is reached (system-dependent effect).

One can proceed as follows: Systems of increasing sizes 
 are successively
considered, and the correlations are computed along $u$- 
(parallel to a square edge considered as the free surface) and $v$-axis (perpendicular
to this edge). The order parameter correlation 
function for example is supposed to obey a scaling
form which reproduces the expected power-law behavior in the thermodynamic 
limit:
\begin{equation}
  G_\perp^{\rm sq}(v)=\frac{1}{ v^{x_\sigma^b+x_\sigma^1}}
  f_\perp^{\rm sq}\left(\frac{v}{N}\right),
  \label{eq: scalG1}
\end{equation}
\begin{equation}
  G_\parallel^{\rm sq}(u)=\frac{1}{ u^{2x_\sigma^1}}
  f_\parallel^{\rm sq}\left(\frac{u}{N}\right),
  \label{eq: scalG2}
\end{equation}
where $x_\sigma^b$ and $x_\sigma^1$ are the bulk and surface 
order parameter scaling dimensions, respectively. The scaling functions have to satisfy  
asymptotic expansions including corrections to scaling due to the limitations
mentioned above, e.g.
$f_\perp^{\rm sq}\left(\frac{v}{N}\right)\sim 1+{\rm const}\times
\left(\frac{v}{N}\right)^\mu+\dots$ in the boundary region $v\to N$.

Equations~(\ref{eq: scalG1}) and (\ref{eq: scalG2}) are not very useful for the determination of critical
exponents, since the scaling regime $v\to N$ is perturbed by the correction 
terms which have a large amplitude, resulting from the significance
of finite-size corrections.  
Nevertheless, conformal invariance supplies an easy way to take into account 
explicitly 
shape effects in two-dimensional systems,  and thus provides a refined procedure 
for the determination of
the exponents. In pure systems, 
density profiles, correlations and local properties have been investigated in
various geometries (surfaces~\cite{cardy84b,gomperwagner85,burkhardtguim87},
corners~\cite{barberpeschelpearce84,daviespeschel91,karevskilajkoturban97}, 
strips~\cite{burkhardteisenriegler86,burkhardtxue91,turbanigloi97} or parabolic 
shapes~\cite{peschelturbanigloi91,turbanberche93b,daviespeschel93,blawidpeschel94,berchedebierreeckle94,kaiserturban94}, 
for a review,
see Ref.~\cite{igloipeschelturban93}), as well as the moments of
the magnetization~\cite{burkhardtderrida85} and structure 
factors~\cite{klebanetal86} have been calculated in square systems.

In the following, we shall consider a square system with free or fixed boundary 
conditions
on all the edges. Using conformal invariance techniques~\cite{christehenkel93}, 
the 
Schwarz-Christoffel mapping enables us to
calculate the surface-bulk correlation fonction inside the square.
The mapping of the complex half-plane $z=x+{\rm i}y$, ${\rm Im}\  z>0$,
 inside a square $\zeta=u+{\rm i}v$, $-N/2\le {\rm Re}\ \zeta\le N/2$, 
 $0\le{\rm Im}\ \zeta\le N$, is realized by the 
 conformal transformation~\cite{lavrentievchabat}
\begin{equation}
  \frac{{\rm d}\zeta}{{\rm d}z}=\frac{C}{\sqrt{(1-z^2)(1-k^2z^2)}}.
  \label{eq: SC1}
\end{equation}
Since $\zeta=N/2$ and $\zeta=N/2+{\rm i}N$ are mapped onto $z=1$ and
$z=1/k$ ($0<k<1$), respectively, the constant $C$ is related to the size of 
the square
\begin{eqnarray}
	N/2C&=&{\rm K}(k)\equiv{\rm K},\nonumber\\ 
	N/C&=&{\rm K}(k')\equiv{\rm K'},
	\label{parameters}
\end{eqnarray} 
where 
$k'=\sqrt{1-k^2}$ and ${\rm K}(k)$ is the complete elliptic integral of 
the first kind. The modulus $k$ also follows from these equations. It is 
given by~\cite{lavrentievchabat}
\begin{equation}
  k=4\left(\frac{\sum_{p=0}^\infty q^{(p+1/2)^2}}{1+2
  \sum_{p=1}^\infty q^{p^2}}\right)^2,\quad q={\rm e}^{-2\pi}.
  \label{eq: k}
\end{equation}
The complete transformation is finally written
\begin{equation}
  \zeta=\frac{N}{{2\rm K}}{\rm F}(z,k)=\frac{N}{{\rm K'}}{\rm F}(z,k),  
  \label{eq: SCdef1}
\end{equation}
\begin{equation}
  z=
  \sn \frac{{\rm K'}\zeta}{N}
  \equiv \sn \left(\frac{{\rm K'}\zeta}{N},k\right),
  \label{eq: SCdef2}
\end{equation}
where ${\rm F}(z,k)$ is the elliptic integral of the first kind and $\sn (\zeta,k)$
the Jacobian elliptic sine~\cite{abramowitz}. 

\subsection{Correlation functions}

The two-point correlation 
function of a conformally invariant system
can now be 
obtained in the $\zeta-$geometry in terms of its counterpart in the semi-infinite
system ($z-$geometry):
\begin{eqnarray}
  G(\zeta_1,\zeta)&\equiv&\langle\sigma(\zeta_1)\sigma(\zeta)
  \rangle_{\rm sq}\nonumber\\
  &=&
  \mid\zeta'(z_1)\mid^{-x_\sigma^b}\mid\zeta'(z)\mid^{-x_\sigma^b}  
  \langle \sigma(z_1)\sigma(z)\rangle_{{\rm hp}},
  \label{eq: TrCor}
\end{eqnarray}
where the correlation function in the half-plane (hp) geometry is known to take the 
form~\cite{cardy84b}
\begin{eqnarray} 
  G(z_1,z)&\equiv&\langle \sigma(z_1)\sigma(z)\rangle_{{\rm hp}}\nonumber\\
  &=&
  {\rm const}\times
  (y_1y)^{-x_\sigma^b}\psi(\omega),
  \label{eq: Corhp}
\end{eqnarray}
where the dependence on $\omega=\frac{y_1y}
  {\mid z_1-z\mid^2}$ of the universal scaling function $\psi$ is constrained
by the special conformal transformation and its asymptotic behavior,
$\psi(\omega)\sim\omega^{x_\sigma^1}$, in the limit $y_1=\order(1)$, $y\gg 1$,
 is implied by scaling. 

Equations.~(\ref{eq: Corhp})  and (\ref{eq: TrCor}), applied in the 
random situation, lead to the correlations between 
$\zeta_1={\rm i}$, close 
to a side of the square, and
any point inside it, as
follows:
\begin{equation}
  [\langle{  G}_\sigma(\zeta)\rangle]_{\rm av}={\rm const}\times 
  \{\mid \zeta'(z)\mid{\rm Im} \left(z(\zeta)\right) \}^{-x_\sigma^b}
  \psi(\omega).
  \label{eq: 10}
\end{equation}
Taking the logarithm of both sides, the bulk critical exponent $x_\sigma^b$ 
can thus be deduced from a  linear fit along $\omega={\rm const}$ 
curves in the square:
\begin{equation}
  \ln [\langle{  G}_\sigma(\zeta)\rangle]_{\rm av}={\rm const'}-x_\sigma^b\ln \kappa(\zeta)+
  \ln\psi(\omega),
  \label{eq: 11}
\end{equation}
with
\begin{equation}
	\kappa(\zeta)\equiv{{\rm Im}(z(\zeta))} {\left|\left[1-
  z^2(\zeta)\right]\left[1-k^2
  z^2(\zeta)\right]\right|^{-1/2}}.	
  \label{eq:kappa}
\end{equation}

We will now discuss the results of MC simulations performed with 
the Swendsen-Wang 
cluster algorithm~\cite{swendsenwang87} for systems of size $101\times 101$
with canonical disorder. 
The details concerning the choice of the parameters for the simulations
(number of MC iterations, \dots) are given in Appendix~\ref{sec:appendixB}. 
Average over disorder is performed
over $N_{\rm rdm}=3000$ samples. All the MC simulations are done at the optimal
disorder amplitude $r^\star(Q)$ determined in the strip geometry.

\begin{figure}
	\epsfxsize=8cm
	\begin{center}
	\mbox{\epsfbox{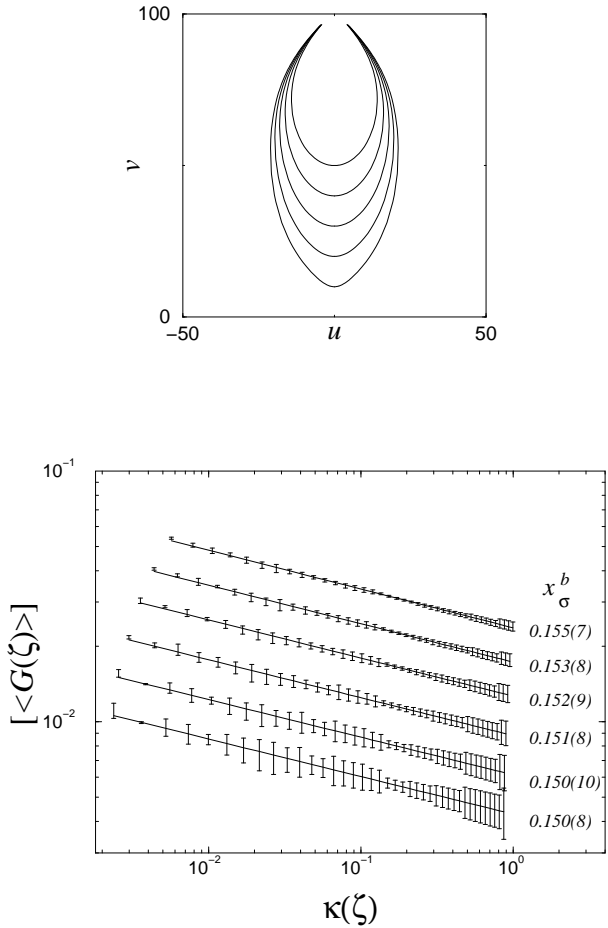}}
	\end{center}
	\caption{Rescaled correlation function along six $\omega={\rm const}$ curves in 
	the square (shown in the upper part). 
	These curves are approximated by linear expansions in the
	neighborhood of the discrete lattice sites, which explains the
	variations on the sizes of error bars ($Q=8$, $r=10$).}
	\label{fig-7}  
\end{figure}

Eq.~(\ref{eq: 11}) is used in Fig.~\ref{fig-7} to extract the bulk
magnetization scaling dimension at $Q=8$. Consistent values are obtained for different
fixed values of the parameter $\omega$. Averaging the 
results at different $\omega$'s, one obtains 
\begin{equation}
x_\sigma^b(8)=0.152\pm 0.003,
\label{eq-x_corr8}
\end{equation}
corresponding to an error of $2\%$.
 
\begin{figure}
	\epsfxsize=5cm
	\begin{center}
	\mbox{\epsfbox{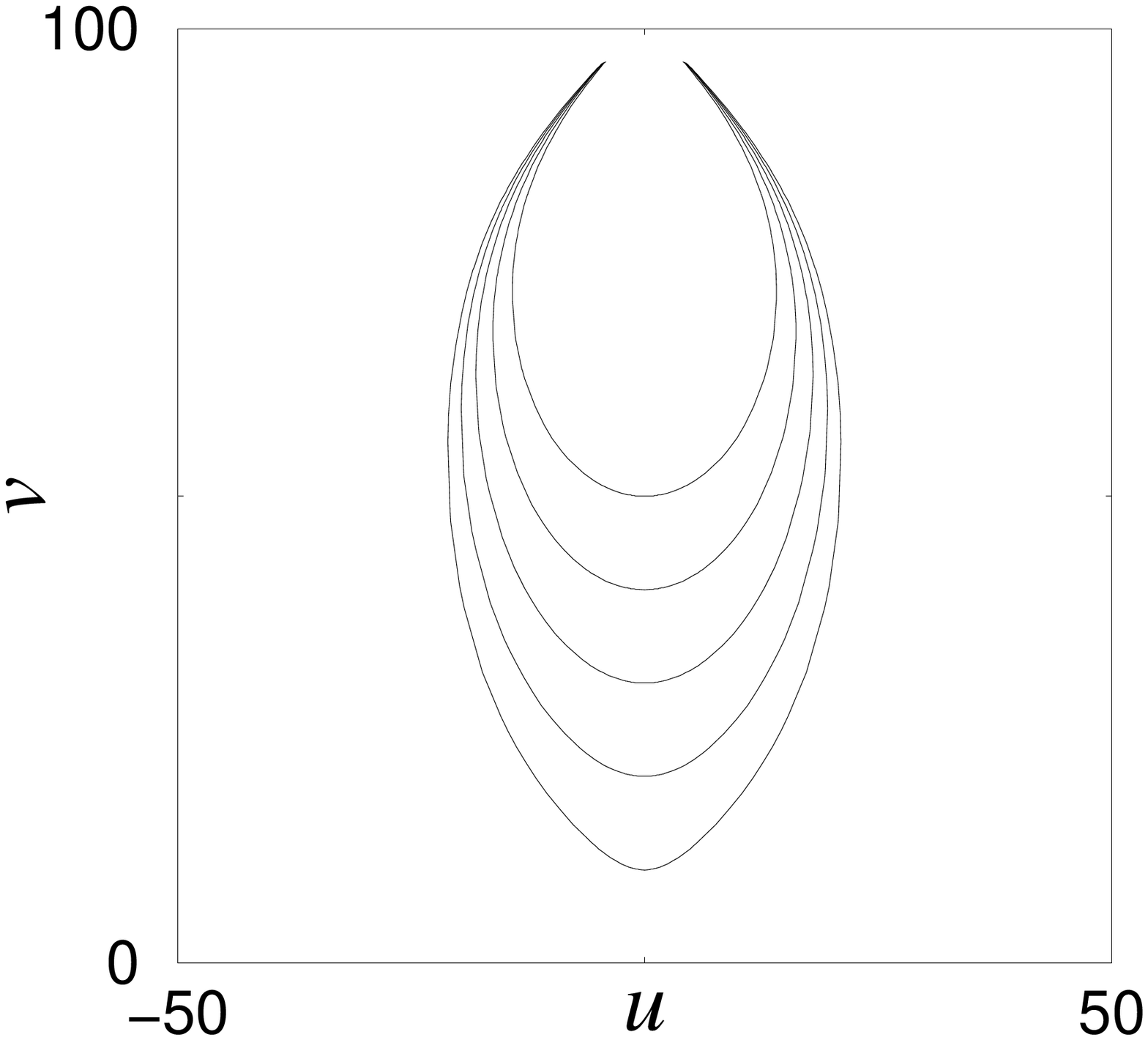}}
	\end{center}
	\end{figure}
\vglue -0.cm 

One should nevertheless mention that the uncertainty on this result is
underestimated, since neither the fluctuations due to randomness, nor
the influence of a variation of $r$ around the optimal value has  
been taken into account explicitly. This is intentional, since such studies
would require intensive computational efforts and would be less accurate that
the next method to be presented.

\subsection{Density profiles}

\begin{figure}
	\epsfxsize=7cm
	\begin{center}
	\mbox{\epsfbox{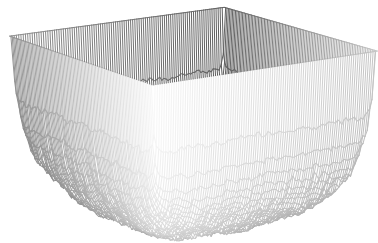}}
	\end{center}\vskip 0.cm
	\caption{Density profiles inside the 
	square averaged for 3000 
	disorder realizations ($Q=8$, $r=10$). 
	}
	\label{3dprofile}  
\end{figure}

Owing to the unknown scaling function $\psi(\omega)$, the determination of
the bulk  critical exponent from the behavior of the correlation function is 
not extremely accurate. Furthemore, since a few points are used for the fits
along $\omega={\rm const.}$ curves, this introduces a poor statistics. 
It can nevertheless be improved if one considers the magnetization
profile inside a square with fixed boundary conditions. Since it is a one-point function,
its decay from the distance to the surface in the semi-infinite geometry is 
fixed, up to a constant prefactor 
\begin{equation}
	[\langle\sigma(z)\rangle_{\rm hp}]_{\rm av}\sim y^{-x_\sigma^b}.
	\label{prof}
\end{equation} 
The local order parameter is 
defined, according to Ref.~\cite{challalandaubinder86}, as the probability for
the spin at site $\zeta$ in the square, to belong to the majority 
orientation (Fig.~\ref{3dprofile}).

\begin{figure}
	\epsfxsize=9cm
	\begin{center}
	\mbox{\epsfbox{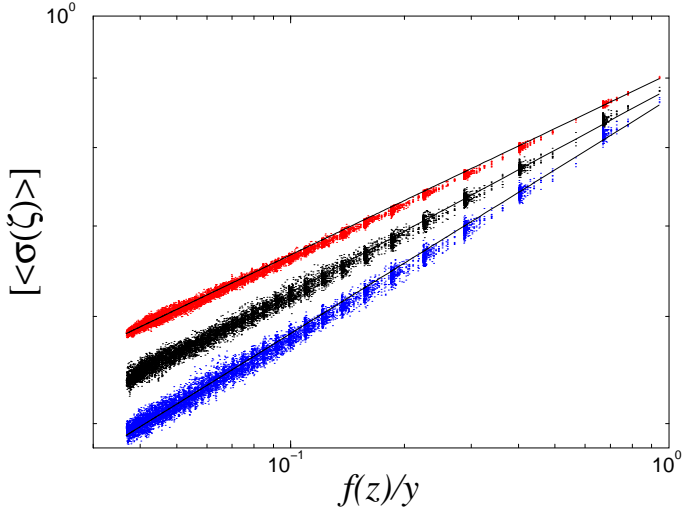}}
	\end{center}\vskip 0.cm
	\caption{Rescaled magnetization profile inside the 
	square for 5000 
	disorder realizations ($Q=3$, 8 and 64 from top to bottom). 
	The power law fits are over $100^2$ data points.}
	\label{fig-8}  
\end{figure}

The Schwarz-Christoffel mapping leads to the following expression
for the average profile in
the square geometry:
\begin{equation}
	[{\langle\sigma (\zeta)\rangle}]_{\rm av}=
	{\rm const}\times\left(\frac{
	\sqrt{|1-z^2(\zeta)|\cdot
	|1-k^2z^2(\zeta)|}
	}{
	{\rm Im}(z(\zeta))}\right)^{x^b_\sigma}.
	\label{eq-prof_conf}
\end{equation}
This expression, of the form $[{\langle\sigma (\zeta)\rangle}]_{\rm av}=
[f(z)/y]^{x^b_\sigma}$, holds for any point inside the square. It 
allows an accurate determination of the 
critical exponent, since the $N^2$ lattice points enter the power-law 
fit (Fig.~\ref{fig-8}).
Although this technique is more precise than the previous one, 
one has to take care to different
sources of error.
It is indeed again necessary to consider the influence of the number of
disorder configurations which are used to get the average magnetization, as well
as the effect of a variation of the disorder amplitude around the optimal
value. 
We performed $N_{\rm rdm}=5000$ realizations 
of disorder
in five independent runs (see Appendix~\ref{sec:appendixB}), and computed the magnetic exponent 
for each run. 
Averaging the results, it yields the values 
given in Table~\ref{tabxmMC}.


\vbox{
\begin{table}
\caption{Bulk magnetic scaling index  obtained from the magnetization profile
inside the square (5000 realizations of disorder).\label{tabxmMC}}
\begin{tabular}{llll}
	 $Q$ & $r$ & $x_\sigma^b$ & $\Delta x_\sigma^b$   \\
	\hline
	$\vphantom{10^{2^2}}$3 & 5 & 0.13357 & $3\times 10^{-5}$ \\
	4 & 7 & 0.13815 & $4\times 10^{-5}$  \\
	5 & 7 & 0.14302 & $4\times 10^{-5}$  \\
	6 & 8 & 0.14621 & $5\times 10^{-5}$  \\
	8 & 10 & 0.15031 & $5\times 10^{-5}$  \\
	15 & 10 & 0.15984 & $6\times 10^{-5}$  \\
	64 & 12 & 0.17299 & $6\times 10^{-5}$  \\
\end{tabular}
\end{table}
}

\subsection{Boundary critical behavior}

The surface scaling dimension can be obtained once the bulk exponent is known.
From standard scaling, the asymptotic behavior of the two-point correlation 
function, when $y_1=\order(1)$, $y\gg 1$ is expected to
involve both bulk and surface dimensions: 
\begin{equation}
	[\langle G_\sigma(y-y_1)\rangle ]_{\rm av}\sim 
	y^{-(x_\sigma^b+x_\sigma^1)}.
	\label{corr-surf}
\end{equation} 

A power law behavior thus follows for the universal scaling function defined
in Eq.~(\ref{eq: Corhp}):
\begin{eqnarray}
\psi(\omega)&\equiv& [\langle G_\sigma(\zeta)\rangle]_{\rm av}\times
\{\mid \zeta'(z)\mid{\rm Im} \left(z(\zeta)\right) \}^{x_\sigma^b}
\nonumber\\
&\sim&
\omega^{x_\sigma^1},\qquad \omega\to 0.
\label{powerlawpsi}
\end{eqnarray}

\begin{figure}
	\epsfxsize=8cm
	\begin{center}
	\mbox{\epsfbox{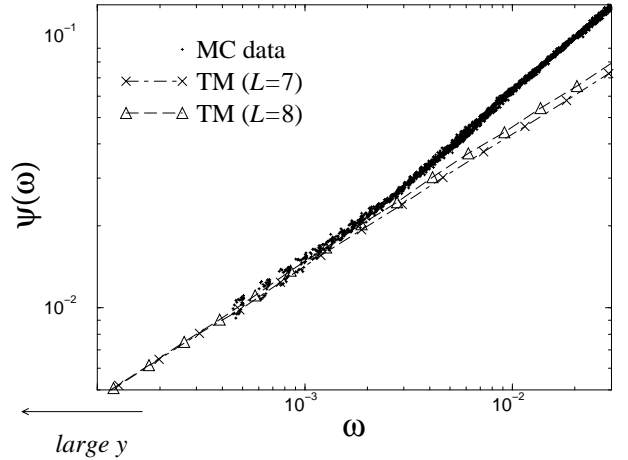}}
	\end{center}
	\caption{Large distance behavior of the universal scaling function
	($Q=8$, $r=10$) leading to the surface scaling index. The fit has been
	shifted for clarity.}
	\label{psi_omega}  
\end{figure}

A log-log plot of Eq.~(\ref{powerlawpsi}) is shown on 
Fig.~\ref{psi_omega}, where the TM results are also presented for comparison. 
The result 
for the surface scaling index is less accurate than in the case of the bulk, 
but
the estimation $x_\sigma^1(8)\simeq 0.47(3)$ is in agreement with the value
that we obtained previously by FSS techniques in Ref.~\cite{chatelainberche98}.
It also agrees with the TM results  which give
$x_\sigma^1(8)\simeq 0.48(2)$ for $L=7$ and $x_\sigma^1(8)\simeq 0.50(2)$
for $L=8$.

If the leading singularity ($x_\sigma^1$) is found to be the same using the two
techniques, we
 note that the corrections to 
 scaling are very different, as it appears in the deviation between
 the curves as $\omega$ increases. This can be the result of the ensemble
 average procedure which is not identical in the two approaches (grand canonical
 for the TM technique and canonical disorder for the MC simulations). The same
 type of sensitivity to the ensemble average was reported recently by Wiseman and 
 Domany~\cite{wisemandomany98}.


\section{Conclusion}\label{sec:sec4}

In this paper, we have investigated the magnetic critical properties of disordered
$Q-$state Potts ferromagnets for a wide range of $Q-$values. 
These models lead to second-order phase transitions which
are particularly interesting, since they belong to new universality classes.
The accurate determination of critical indices is a preliminary step towards a
deeper understanding of these universality classes.
Although universality is expected with respect to the disorder amplitude $r$, previous
works on finite systems have shown that the numerical results are very 
sensitive to the choice of
this disorder amplitude. This sensitivity is attributed to crossover effects
due to the pure model ($r\to 1$) and percolation ($r\to\infty$) 
unstable fixed points. The behavior of the effective central charge
as a function of $r$ can fortunately be exploited to locate the optimal regime
of disorder.
One should mention that in our previous studies, this extreme sensitivity of
the numerical estimates of critical exponents was not well understood, resulting
in an underestimation of uncertainties. We
tried to present here a carefull analysis leading to reliable error bars.
This uncertainty is mainly due to the non self-averaging
behavior of correlation functions. In the strip geometry, 
the number of samples being already important, better
estimates would not be easy to obtain,
 whilst in the MC simulations, improvements
could be supplied by increasing the number of realizations of disorder. 

The
conformal mapping inside the square seems very efficient compared to standard
FSS studies,  one lattice size being needed only. The accuracy
is furthermore substantially improved, since

i) the finite-size corrections
are essentially included in the conformal mapping,

ii) all the lattice points enter the fit of the density profiles.
 
\begin{figure}
	\epsfxsize=8cm
	\begin{center}
	\mbox{\epsfbox{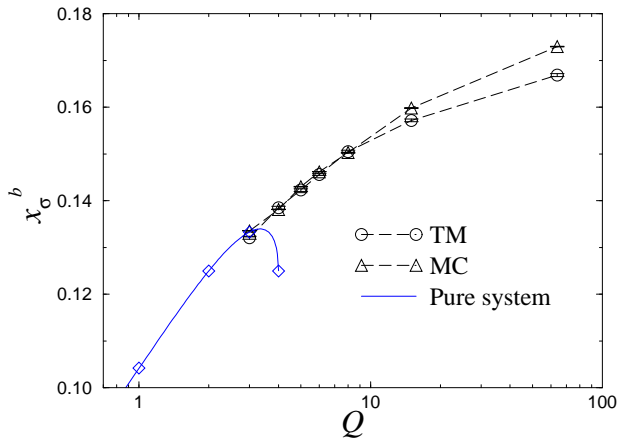}}
	\end{center}
	\caption{$Q$--dependence of the bulk magnetic scaling dimension
	in the RBPM compared to the pure model value for 
	$Q\le 4$.}
	\label{x_vs_Q}  
\end{figure}

A summary of our results, compared to other independent determinations 
of the magnetic scaling index, is given in Table~\ref{tabxm}, and the dependence on $Q$ is shown 
in Fig.~\ref{x_vs_Q}. The pure model value for $Q\le 4$ is shown for
comparison~\cite{baxter82}. Both FSS and
conformal invariance results are presented. The two techniques used in this work
are in agreement with each other, as well as with previous studies 
at the same disorder amplitude, at least as long as the number of states $Q$ is
not too large.
When the ratio $r$ is very different, disagreement with other studies 
which are likely 
due to crossover effects occurs. On the other hand, when the number of states 
is large, $Q>15$,
there appears discrepancies between the two techniques used here. Whilst the
second method (square geometry, 5000 realizations) seems to be the most 
accurate, we are more confident in the
first one (strip geometry, 80000 realizations): If the
number of disorder realisations is too small, the average
behaviour will indeed give an exponent closer to the typical one, 
and thus too large. MC simulations are furthermore known to be less efficient
when $Q$ increases, since the autocorrelation time increases also, requiring
larger numbers of thermalisation iterations.

 We also note  that the leading singularity of the magnetization does not depend,
 up to the precision of our results, 
on the type of disorder considered (GCD or CD).

\end{multicols}
\vbox{\widetext
\begin{table}
\caption{Extrapolation of the bulk magnetic scaling dimension $x_\sigma^b$ 
in the thermodynamic
limit for the different 
values of $Q$. The first two columns recall previous FSS results obtained by MC 
simulations (in which the accuracy had been overestimated, since the
influence of the disorder amplitude was not well understood, at least
in which concerns our own studies). The data in the four remaining columns were deduced from conformal
invariance. The quantity that was studied is indicated in the table as well as
the geometry and the numerical technique. The results presented in this work are
written in bold face. The table notes recall the parameters used for each result,
especially the values of disorder amplitude which are known to have strong
influence on the exponent.
\label{tabxm}}
\begin{tabular}{lllllll}
\multicolumn{1}{c}{}&\multicolumn{2}{c}{\hfill FSS(MC)\hfill\vline}&\multicolumn{4}{c}{Conformal Invariance}\\
\multicolumn{1}{c}{}&\multicolumn{2}{c}{\hfill square\hfill\vline}&\multicolumn{2}{c}{strip}&\multicolumn{2}{c}{square}\\
&SW & W\hfill\vline & TM & TM & SW & SW \\
$Q$\hfill\vline & $[\langle M_b\rangle]$ & $[\langle M_b\rangle]$\hfill\vline & $[\langle G(u)\rangle]$ & $[\langle G (u)\rangle]$ 
& $[\langle   G (\zeta)\rangle]$ & $[\langle \sigma (\zeta)\rangle]$ \\
\tableline
3\hfill\vline &  & 0.1337(7)$^{\rm a}$\ \hfill\vline & 0.1347(1)$^{\rm b}$ &\bf 0.1321(3)$^{\rm c}$ &  &\bf 0.13357(3)$^{\rm d}$  \\  
4\hfill\vline & 0.145(3)$^{\rm e}$  & 0.139$^{\rm f}$\hfill\vline & 0.1396(5)$^{\rm b}$ &\bf 0.1385(3)$^{\rm c}$  &  & \bf 0.13815(4)$^{\rm d}$  \\  
5\hfill\vline &  &\hfill\vline  & 0.1413(10)$^{\rm b}$ &\bf 0.1423(3)$^{\rm c}$  &  & \bf 0.14302(4)$^{\rm d}$  \\  
6\hfill\vline &  &\hfill\vline  & 0.1423(9)$^{\rm b}$ &\bf 0.1456(3)$^{\rm c}$  &  & \bf 0.14621(5)$^{\rm d}$  \\  
8\hfill\vline & 0.118(2)$^{\rm g}$ & 0.153(1)$^{\rm h}$\hfill\vline & 0.1415(36)$^{\rm b}$ &  &  &   \\  
8\hfill\vline & 0.153(3)$^{\rm i}$ &0.151(4)$^{\rm h}$\hfill\vline  & 0.1496(9)$^{\rm j}$ & \bf 0.1505(3)$^{\rm c}$ & 0.152(3)$^{\rm k}$ &\bf 0.15031(5)$^{\rm d}$   \\
15\ \hfill\vline&  &\hfill\vline  &  &\bf 0.1572(3)$^{\rm c}$  &  & \bf 0.15984(6)$^{\rm d}$  \\  
64\hfill\vline&  & 0.185(5)$^{\rm h}\hfill\vline $  &  &\bf 0.1669(3)$^{\rm c}$  &  & \bf 0.17299(6)$^{\rm d}$  \\  
\end{tabular}
\tablenotetext[1]{MC simulations (Wolff algorithm, $\sim 10^5$ samples, $r=10$, GCD) from 
Ref.~\protect\(\cite{picco96}\protect\).} 
\tablenotetext[2]{TM calculations ($L=1-7$, $10^2$ samples, $r=2$, GCD) from 
Refs.~\protect\(\cite{cardyjacobsen97,jacobsencardy98}\protect\).}
\tablenotetext[3]{TM calculations ($L=2-8$, $80\times 10^3$ samples, the values of disorder amplitude for $Q=3$, 4, 5, 6, 8,
15 and 64 are $r=5$, 7, 7, 8, 10, 10 and 12, respectively, GCD), this work.}
\tablenotetext[4]{MC simulations (Swendsen-Wang algorithm, $N=101$,
$5\times 10^3$ samples, the values of disorder amplitude for $Q=3$, 4, 5, 6, 8,
15 and 64 are $r=5$, 7, 7, 8, 10, 10 and 12, respectively, CD), this work.}
\tablenotetext[5]{MC simulations (cluster algorithm, $N=256$, 
$\sim 500$ samples, $r=10$, GCD) from 
Ref.~\protect\(\cite{wisemandomany95}\protect\).}
\tablenotetext[6]{MC simulations (Wolff algorithm)  M. Picco, Ref. [52] cited in 
Ref.~\protect\(\cite{jacobsencardy98}\protect\).}
\tablenotetext[7]{MC simulations (Swendsen-Wang algorithm, 
$N\leq 100$, $\sim 30$ samples, $r=2$, restricted CD) from 
Ref.~\protect\(\cite{chenferrenberglandau95}\protect\).}
\tablenotetext[8]{MC simulations (Wolff algorithm, $N\leq 100$ and 500, $\sim 10^5$ samples, $r=10$, GCD) from 
Ref.~\protect\(\cite{picco98}\protect\).} 
\tablenotetext[9]{MC simulations (Swendsen-Wang algorithm, 
$N\leq 100$, $\sim 500$ samples, $r=10$, CD) from 
Ref.~\protect\(\cite{chatelainberche98}\protect\).}
\tablenotetext[10]{TM calculations ($L=2-9$, $40\times 10^3$ samples, $r=10$, GCD) from 
Ref.~\protect\(\cite{chatelainberche98b}\protect\).}
\tablenotetext[11]{MC simulations (Swendsen-Wang algorithm, $N=101$, 
$3\times 10^3$ samples, $r=10$, CD) from 
Ref.~\protect\(\cite{chatelainberche98b}\protect\).}
\end{table}
%
}

\begin{multicols}{2}
\narrowtext

In the case $Q=3$ which was already considered by different authors, there 
exists
a perturbative result (renormalization group approach for the 
perturbative series around the pure model
conformal field theory):
\begin{equation}
	x_\sigma^b=\frac{2}{15}+0.00132\simeq 0.13465
	\label{exp-Q=3}
\end{equation}
This result was confirmed numerically by Picco~\cite{picco96} 
and Cardy and Jacobsen~\cite{cardyjacobsen97}. In this work, we obtain a value
which is slightly too small.
We
nevertheless note that the two values, at $r=5$ and $r=6$ are in 
perfect agreement
(see Appendix~\ref{sec:appendixB}).

We eventually mention a recent work of Olson and Young~\cite{olsonyoung99}
who performed a MC study of the multiscaling properties of the correlation functions for
different values of $Q$. They used a different self-dual probability
distribution of the couplings, and obtained slightly different results
(e.g. $x_\sigma^b=0.161(3)$ at $Q=8$).

\acknowledgments
  We thank L. Turban for  critically reading  the manuscript 
  and J.L. Jacobsen for suggesting us to control the stability of the
  exponents around the optimal disorder amplitude. We are also indebted to M. Picco
  who drew our attention on refined analysis of the data which reduces the error
  bars.
 
  Due to the disorder average, the numerical study of disordered 
  systems is particularly suitable to
  parallel computing.
The computations presented here were performed on the {\it SP2} at
 the CNUSC in Montpellier under projects No. C981009 and C990011,  and the 
 {\it Power Challenge Array} at the CCH in Nancy. The Laboratoire de Physique des Mat\'eriaux is Unit\'e Mixte de Recherche  
 CNRS No. 7556.

\end{multicols}

\widetext
\begin{appendix}
\section{Evaluation of errors in the Transfer Matrix calculations}
\label{sec:appendixA}
\begin{multicols}{2}

In spite of the large number of disorder
realizations, the correlation functions along the strip display an 
important dispersion
but the resulting values for the critical exponents are  extremely accurate.
In order to obtain a correct estimation of the errors on the  
magnetic scaling index, we studied the influence of the cutoff parameter 
$\epsilon$. For $\epsilon\simeq 10^{-1}$, a few points are taken into account 
only and the short distance behavior of the correlation function is observed. On the
other hand, with $\epsilon\simeq 10^{-6}$, all the data points in the range $u=5-100$
are taken into account in the fit, giving a greater weight to the 
long-distance behavior. Clearly, one has to find 
a compromise between the two approaches. 
Fortunately a variation of the cutoff parameter does not  
affect  the value of the extrapolated
exponent which remains very stable
as shown in 
Table~\ref{tab-error}.

\end{multicols}\vbox{
\widetext
\begin{table}
\caption{Bulk magnetic scaling index (after extrapolation in the 
thermodynamic limit) obtained from the decay of the correlation
function along the strip with different values of the cutoff $\epsilon$.
\label{tab-error}}
\begin{tabular}{l|llllllll}
&\multicolumn{8}{c}{Effective exponent at $Q=3$, $r=5$}\\
\tableline
$\epsilon$ & 0.125 $\times 10^{-1}$ & 0.312 $\times 10^{-2}$ & 0.781 $\times 10^{-3}$ & 0.195 $\times 10^{-3}$ & 0.488 $\times 10^{-4}$ & 0.122 $\times 10^{-4}$ & 0.305 $\times 10^{-5}$  &  0.763 $\times 10^{-6}$  \\
$x_\sigma^b$ & 0.13207 & 0.13209 & 0.13209 & 0.13208 & 0.13209 & 0.13209 & 0.13209 & 0.13209 \\
$\Delta x_\sigma^b$ & 5.3$\times 10^{-4}$ & 3.8$\times 10^{-4}$ & 3.3$\times 10^{-4}$ & 3.1$\times 10^{-4}$ & 3.1$\times 10^{-4}$ & 3.0$\times 10^{-4}$ & 3.0$\times 10^{-4}$  & 3.0$\times 10^{-4}$ \\
\tableline
\tableline
&\multicolumn{8}{c}{Effective exponent at $Q=8$, $r=10$}\\
\tableline
$\epsilon$ & 0.125 $\times 10^{-1}$ & 0.312 $\times 10^{-2}$ & 0.781 $\times 10^{-3}$ & 0.195 $\times 10^{-3}$ & 0.488 $\times 10^{-4}$ & 0.122 $\times 10^{-4}$ & 0.305 $\times 10^{-5}$  &  0.763 $\times 10^{-6}$  \\
$x_\sigma^b$ &0.15014  & 0.15032 & 0.15047 &0.15050  & 0.15050 & 0.15053 & 0.15054 & 0.15054 \\
$\Delta x_\sigma^b$ &4.7$\times 10^{-4}$  & 3.4$\times 10^{-4}$ & 2.9$\times 10^{-4}$ & 2.7$\times 10^{-4}$  & 2.7$\times 10^{-4}$ & 2.6$\times 10^{-4}$ & 2.6$\times 10^{-4}$ & 2.6$\times 10^{-4}$ \\
\tableline
\tableline
&\multicolumn{8}{c}{Effective exponent at $Q=8$, $r=11$}\\
\tableline
$\epsilon$ & 0.125 $\times 10^{-1}$ & 0.312 $\times 10^{-2}$ & 0.781 $\times 10^{-3}$ & 0.195 $\times 10^{-3}$ & 0.488 $\times 10^{-4}$ & 0.122 $\times 10^{-4}$ & 0.305 $\times 10^{-5}$  &  0.763 $\times 10^{-6}$  \\
$x_\sigma^b$ &0.15040  & 0.15056 & 0.15072 &0.15071  & 0.15074 & 0.15077 & 0.15078 & 0.15078 \\
$\Delta x_\sigma^b$ &4.7$\times 10^{-4}$  & 3.3$\times 10^{-4}$ & 2.9$\times 10^{-4}$ & 2.7$\times 10^{-4}$  & 2.6$\times 10^{-4}$ & 2.6$\times 10^{-4}$ & 2.6$\times 10^{-4}$ & 2.6$\times 10^{-4}$ \\
\tableline
\tableline
&\multicolumn{8}{c}{Effective exponent at $Q=64$, $r=12$}\\
\tableline
$\epsilon$ & 0.125 $\times 10^{-1}$ & 0.312 $\times 10^{-2}$ & 0.781 $\times 10^{-3}$ & 0.195 $\times 10^{-3}$ & 0.488 $\times 10^{-4}$ & 0.122 $\times 10^{-4}$ & 0.305 $\times 10^{-5}$  &  0.763 $\times 10^{-6}$  \\
$x_\sigma^b$ & 0.1663 & 0.1663 & 0.1667 & 0.1668 & 0.1668 & 0.1669 & 0.1670 & 0.1671 \\
$\Delta x_\sigma^b$ & 6$\times 10^{-4}$ & 4$\times 10^{-4}$  &3$\times 10^{-4}$ & 3$\times 10^{-4}$ & 3$\times 10^{-4}$ & 3$\times 10^{-4}$ & 3$\times 10^{-4}$ & 3$\times 10^{-4}$ \\
\end{tabular}
\end{table}}
%
\begin{multicols}{2}

Another contribution to the error should come from the choice of the
disorder amplitude. To study this effect, we considered a variation of $r$
 close to the optimal value. It leads to a result which is inside 
 the error bars of the previous one, as shown in the case $Q=8$ in
 Table~\ref{tab-error}.
 The  uncertainty in the range $\epsilon=10^{-4}-10^{-6}$ is of 
 the same order of magnitude than 
the
fluctuations between the data obtained with different values of $\epsilon$
and $r$, so
we eventually consider as a definitive result the fit with this cutoff
 value.

\section{Details of the Monte Carlo simulations}
\label{sec:appendixB}

In random systems, in addition to the usual MC error, the random-bond
fluctuations introduce another source of statistical error. For any physical quantity
$X$, the total error is given by
\begin{equation}
(\delta X)^2=\frac{\sigma_{\rm rdm}^2}{N_{\rm rdm}}+\frac{\sigma_T^2(1+2\tau_X)}{N_{\rm rdm}N_{\rm MC}}
\label{eq-error}
\end{equation}
where the first term is due to the disorder fluctuations, whilst the
second one describes the fluctuations during the MC iterations. This latter
term corresponds
to the standard deviation of independent random variables, corrected by the
autocorrelation time to take into account the correlations between the 
successive data. In these expressions,
$N_{\rm MC}$ is the number of MC iterations, measured in MC steps (MCS), 
realized for the measurements of the physical
quantities for each disorder realization,
$N_{\rm rdm}$ is the number of disorder realizations and $\tau_X$ is the 
autocorrelation time for the quantity $X$ (the definition of $\tau_X$ 
sometimes absorbs the factor 1 describing uncorrelated variables).
The variances $\sigma_T$ and $\sigma_{\rm rdm}$ respectively measure the deviation
due to thermal fluctuations for a given sample and the deviation from the
exact value within the ensemble of disorder configurations.

Both variances are of the same order of magnitude. The leading source
of error thus comes from the disorder average and a large number of samples is needed
in order to get accurate results. 
In our simulations we used the Swendsen-Wang 
cluster algorithm~\cite{swendsenwang87} for systems of size $101\times 101$. 
The autocorrelation time for the total magnetization
is $\tau_\sigma\simeq 35$~MCS. The preliminary 5000~MCS have been 
discarded for thermalisation (better that $10^2\times\tau_\sigma$), and 
$N_{\rm MC}=10^4$~MCS
were done to compute the physical quantities. 
Average over disorder is performed
over $N_{\rm rdm}=5000$ samples. From preliminary runs over 1000 samples, we deduced the
standard deviations $\sigma^2_{\rm rdm}\simeq 0.93$ and $\sigma^2_{\rm MC}\simeq 0.13$.
The order of magnitude of the two contributions to the error is thus
\begin{eqnarray}
	\delta\sigma_{\rm MC} &\simeq& \sqrt{
	\frac{\sigma_T^2(1+2\tau_\sigma)}
	{N_{\rm rdm}N_{\rm MC}}
	}
	\simeq 6\times 10^{-4},\nonumber\\	
	\delta\sigma_{\rm rdm} &\simeq& \sqrt{
	\frac{\sigma_{\rm rdm}^2}
	{N_{\rm rdm}}
	}
	\simeq 1.36\times 10^{-2},
	\label{error-MC-rdm}
\end{eqnarray}
for a point in the middle of the square. Due to the fixed boundary conditions,
close to the edges of the square
the fluctuations are reduced. 
These values confirm the  significance of the disorder contribution 
($\frac{\delta\sigma_{\rm MC}}{[\langle\sigma\rangle]_{\rm av}}\simeq 0.08\%$ and
$\frac{\delta\sigma_{\rm rdm}}{[\langle\sigma\rangle]_{\rm av}}\simeq 2\%$). 
  The values of the exponent $x_\sigma^b$ for different values
of $Q$ and $r$ are given in Table~\ref{exposantMC}.

\end{multicols}
\widetext
\begin{table}
\caption{Bulk magnetic scaling index  obtained from the profile of the order parameter
inside a square with fixed boundary conditions for five independent runs 
(1000 configurations of disorder for each run). The final result obtained with 
5000 configurations of disorder is
given in the column called average.
\label{exposantMC}}
\begin{tabular}{l|lllll|l}
&\multicolumn{6}{c}{Exponent at $Q=3$, $r=5$}\\
\tableline
& run 1 & run 2 & run 3 & run 4 &  run 5 & average \\
$x_\sigma^b$ & 0.13260 & 0.13384 & 0.13405 & 0.13418 & 0.13323 & 0.13357  \\
$\Delta x_\sigma^b$ & 7$\times 10^{-5}$ & 7$\times 10^{-5}$ & 7$\times 10^{-5}$ & 7$\times 10^{-5}$ & 7$\times 10^{-5}$ & 3$\times 10^{-5}$ \\
\tableline
\tableline
&\multicolumn{6}{c}{Exponent at $Q=3$, $r=6$}\\
\tableline
& run 1 & run 2 & run 3  & run 4 & run 5  & average  \\
$x_\sigma^b$ & 0.13450 & 0.13262 & 0.13307 & 0.13378 & 0.13333 & 0.13345 \\
$\Delta x_\sigma^b$ & 7$\times 10^{-5}$ & 7$\times 10^{-5}$ & 7$\times 10^{-5}$ & 7$\times 10^{-5}$ & 7$\times 10^{-5}$ & 3$\times 10^{-5}$ \\
\tableline
\tableline
&\multicolumn{6}{c}{Exponent at $Q=4$, $r=7$}\\
\tableline
& run 1 & run 2 & run 3  & run 4 & run 5  & average  \\
$x_\sigma^b$ & 0.13886 & 0.13835 & 0.13798 & 0.13703 & 0.13858 & 0.13815 \\
$\Delta x_\sigma^b$ & 8$\times 10^{-5}$ & 8$\times 10^{-5}$ & 8$\times 10^{-5}$ & 8$\times 10^{-5}$ & 8$\times 10^{-5}$ & 4$\times 10^{-5}$ \\
\tableline
\tableline
&\multicolumn{6}{c}{Exponent at $Q=4$, $r=8$}\\
\tableline
& run 1 & run 2 & run 3  & run 4 & run 5  & average  \\
$x_\sigma^b$ & 0.13799 & 0.13794 & 0.13753 & 0.13849 & 0.13798 & 0.1379 \\
$\Delta x_\sigma^b$ & 9$\times 10^{-5}$ & 9$\times 10^{-5}$ & 9$\times 10^{-5}$ & 9$\times 10^{-5}$ & 9$\times 10^{-5}$ & 4$\times 10^{-5}$ \\
\tableline
\tableline
&\multicolumn{6}{c}{Exponent at $Q=8$, $r=10$}\\
\tableline
& run 1 & run 2 & run 3 & run 4 &  run 5  & average\\
$x_\sigma^b$ & 0.1508 & 0.1515 & 0.1501 & 0.1501 & 0.1492 & 0.15031 \\
$\Delta x_\sigma^b$ & 1$\times 10^{-4}$ & 1$\times 10^{-4}$ & 1$\times 10^{-4}$ & 1$\times 10^{-4}$ & 1$\times 10^{-4}$ & 5$\times 10^{-5}$ \\
\tableline
\tableline
&\multicolumn{6}{c}{Exponent at $Q=8$, $r=20$}\\
\tableline
& run 1 & run 2 & run 3 &  &   & average \\
$x_\sigma^b$ & 0.14527 & 0.14506 & 0.14505 &  & & 0.14513  \\
$\Delta x_\sigma^b$ & 6$\times 10^{-5}$ & $6\times 10^{-5}$ &   $6\times 10^{-5}$ &    &  &   $3\times 10^{-5}$  \\
\tableline
\tableline
&\multicolumn{6}{c}{Exponent at $Q=64$, $r=12$}\\
\tableline
& run 1 & run 2 & run 3 & run 4 &  run 5 & average \\
$x_\sigma^b$ & 0.1722 & 0.1733 & 0.1724 & 0.1747 & 0.1725 & 0.17299  \\
$\Delta x_\sigma^b$ & $1\times 10^{-4}$ & $1\times 10^{-4}$ &   $1\times 10^{-4}$ &   $1\times 10^{-4}$ &  $1\times 10^{-4}$ &  $6\times 10^{-5}$ \\
\end{tabular}
\end{table}
%
\begin{multicols}{2}
 
\end{multicols}
 
\end{appendix}
\begin{multicols}{2}


\newcommand{\Name}[1]{\rm  #1,}
\newcommand{\And}{\ and\ }
\newcommand{\Review}[1]{\it  #1\rm}
\newcommand{\Vol}[1]{\bf  #1\rm,}
\newcommand{\Year}[1]{\rm  (#1)}
\newcommand{\Page}[1]{\rm  #1}
\newcommand{\Book}[1]{\it  #1\rm}

\def\JPA{\it J. Phys. A: Math. Gen.}
\def\JPC{\it J. Phys. C: Solid State Phys.}
\def\PRB{\it Phys. Rev. B}
\def\PRE{\it Phys. Rev. E}
\def\PRL{\it Phys. Rev. Lett.}
\def\JSP{\it J. Stat. Phys.}
\def\JPF{\it J. Phys.  I France}
\def\ZPB{\it Z. Phys. B}

\def\paper#1#2#3#4#5{#1, #3 {\bf #4}, \rm #5 (#2).}

\end{multicols}

\end{document}